# A Hybrid-Cloud Management Plane for Data Processing Pipelines

Vignesh Babu
University of Illinois at Urbana Champaign

Feng Lu, Haotian Wu, Cameron Moberg
Google Cloud

## Abstract

As organizations increasingly rely on data-driven insights, the ability to run data intensive applications seamlessly across multiple cloud environments becomes critical for tapping into cloud innovations while complying with various security and regulatory requirements. However, big data application development and deployment remain challenging to accomplish in such environments. With the increasing containerization and modernization of big data applications, we argue that a unified control/management plane now makes sense for running these applications in hybrid cloud environments. To this end, we study the problem of building a generic hybrid-cloud management plane to radically simplify managing big data applications. A generic architecture for hybrid-cloud management, called *Titchener*, is proposed in this paper. Titchener comprises of independent and loosely coupled local control planes interacting with a highly available public cloud hosted global management plane. We describe a possible instantiation of Titchener based on Kubernetes and address issues related to global service discovery, network connectivity and access control enforcement. We also validate our proposed designs with a real management plane implementation based on a popular big data workflow orchestration in hybrid-cloud environments.

## 1 Introduction

The data analytics market has been forecast to exceed $64B by 2021 with an annual growth rate of 29% [11]. Big data applications have enabled organizations of all sizes to derive meaningful business insights from massive datasets. Cloud vendors offer big data applications on demand via fully managed analytics services (e.g. SaaS) or enable one to build these applications with great ease (e.g. IaaS). While these offerings have allowed enterprise customers to surpass conventional limitations of scale, performance, and cost-efficiency, the need to run big data applications on-premises persists. There are many legitimate reasons, of which legal, compliance, and security [36] stand out, to run parts of a big data system (e.g. enterprise data warehouse) on private clouds. The need to host parts of big data applications in private clouds is also evidenced by recent studies [8, 9]. While most enterprise customers have heavily invested in public clouds, 95% state that they plan to sustain or increase their investment in private clouds and believe that a hybrid-cloud deployment is likely "the end state...over the long term"[9].

While a plethora of prior works exist on coordination, metadata management and monitoring [29, 33] of distributed data processing tasks in single cloud environments, there is little prior art on seamless extension of these services to run data processing tasks spread across public and private clouds. Not surprisingly, managing hybrid-cloud big data applications is a notorious task for several reasons. Big data applications are fairly monolithic [34], and as a result, enterprise customers often have to make the all-or-nothing deployment decision. The problem is exacerbated by the lack of a common infrastructure abstraction layer, it becomes very complex for application developers to work with low-level layers such as VMs and heterogeneous cloud APIs [28]. Further, network virtualization and inter-cloud connectivity is typically hard to set up, configure, and maintain [20].

Motivated by recent trends in big data applications and cloud service advancements, we present a generic hybrid-cloud management plane that blurs the boundary between public and private cloud. In an attempt to radically simplify managing and running hybrid big data applications, we vouch for a public cloud backed overarching management plane that seamlessly interconnects multiple distributed (identical) control and data planes. This management layer provides a single pane of glass to administrate, deploy, execute, and operationalize big data applications on hybrid cloud environments. We believe our architecture is viable and aligns with the following observations:

- Emergence of a common abstraction layer. The big data community has made tremendous progress in containerizing big data applications and embracing Kubernetes as the common portability layer. For example, Google recently announced the support of running Apache Spark [37] and Flink [22] applications on Kubernetes [12, 10]. Further down the stack, Apache Yunikorn [5] aims to provide a YARN-like experience on Kuberenetes.

- Increased adoption of microservice architecture. While Apache Yarn [34] aims to modularize the Hadoop compute platform, projects like Alluxio [1] and AWS Glue [6] went further to make certain functionalities as stand-alone services (e.g., storage and technical metadata). We believe that customers will be presented with a lot more flexibility when it comes to mix and match applications spanning between public and private clouds.

- Networking support. Cloud vendors continue to push the limits of network virtualization across hybrid-cloud environments. In addition, we show that the current cloud networking support is adequate for running hybrid-cloud big data applications with a restricted yet still useful form of network connectivity.

The rest of this paper is organized as follows. In Section 2, we describe Titchener, a system architecture for management of hybrid-cloud data processing pipelines and elaborate why such a design is a good fit in Section 3. In Section 4, we use this architecture to build a hybrid Kubernetes platform for managing containerized hybrid-cloud data processing pipelines. We use the hybrid Kubernetes platform to provide a hybrid-cloud Apache Airflow service in Section 5. In Sections 6 and 7 we present related work and additional discussions respectively and conclude in Section 8.

## 2 Architecture Design

We envision a management plane which can be seamlessly partitioned across multiple cloud environments, more specifically:

**Distributed and loosely coupled control planes.** Components of the overarching management plane are co-located with distributed control planes. Each deployment consists of: (1) local management plane components and (2) forwarding elements to configure these components and interconnect them with application control planes. We believe such a design choice would allow application-specific control planes to independently scale and still present a unified management plane interface.

**Public cloud hosted master control/management plane.** Public cloud vendors promise high availability, e.g, Google Cloud Platform offers 4-9s SLA[7]. Due to such high availability guarantees, we treat the public cloud as an always-on master to service coordination requests from individual forwarding elements and application control planes (if any).

**Single pane of glass.** To match the user experience in a single cloud environment (e.g., public or private-only cloud), the system interface should hide the existence of multiple control/data planes and provide transparency to end users as a single service. This would greatly reduce the operation and management complexity.

The overall architecture of our envisioned hybrid-cloud management plane (dubbed Titchener) is described here (see Figure 1a):

**(i) Containerization.** To reduce operational complexity associated with application deployment and migration, we recommend the creation of a common infrastructure layer between public and private clouds by using container orchestration platforms to host all system and application components.

**(ii) Unified Client Interface.** It should be possible to use the same API, UI and CLI interface tools in a hybrid deployment. A unified client interface would simplify user authentication, authorization, and job ingestion.

**(iii) Strongly Consistent Overwatch Service.** We recommend the implementation of a strongly consistent overwatch service backed by a cloud-managed RDBMS to handle initialization tasks such as control plane discovery, registration and configuration.

**(iv) Intelligent Job Dispatcher.** This module takes user input, checks with a pre-defined service routing rule, and dispatches the work to a specific control plane.

**(v) Control Agent.** We recommend the creation of a control agent along with each control plane. It serves as a local forwarding element and handles job acceptance, submission to local control plane, execution tracking, health and telemetry reporting.

**(vi) Secure and Private Connectivity.** We recommend the creation of secure and non-interfering channels over the internet to allow data exchange between local and public-cloud hosted (1) control agents and (2) management plane components.

**(vii) Distributed Management Planes.** Control agents push configurations to local management plane components. To each application component, the system looks no different from running in a single public/private-only cloud environment.

In the next section, we elaborate how the above design simplifies management of hybrid-cloud big data pipelines.

## 3 Motivation

Popular data processing frameworks like [37, 2, 4] rely on critical management services such as [29, 33, 3, 13] for global coordination, resource allocation and execution tracking of submitted tasks. Extending such services to a hybrid-cloud environment is challenging because the components of the data plane may be partitioned across multiple clouds. However, due to the following reasons, we believe Titchener can address these challenges.

Titchener calls for a partitioned management plane comprising of localized and loosely-coupled control planes. We anticipate most communication to happen locally with occasional control and management traffic between public and private clouds. Such design decision greatly simplifies network policy config, set-up, and enforcement. In addition, any application-level coupling (e.g., resource allocation) between control and data plane

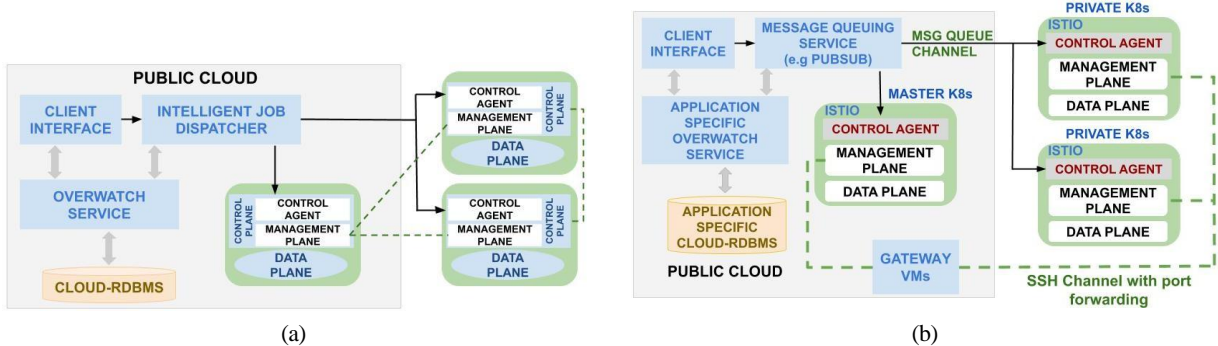

Figure 1: (a) Titchener: hybrid-cloud management plane architecture (b) A Hybrid Kubernetes platform which serves as a run-time fabric for managing containerized data processing pipelines

is preserved due to self-contained localized control and data planes. Given that most data processing jobs are organized as extract-transform-load (ETL) pipelines, Titchener can easily make task-level orchestration and job dispatch decisions depending on policy constraints and available backends (hosted on public or private cloud). The single pane of glass model presents a unified view for operation and policy management, which is made possible by limiting traffic to control and management plane only. Finally, a containerized common infrastructure layer can make it easier to deploy, configure and migrate existing applications across widely heterogeneous environments leading to streamlined application deployment.

So far we have argued why Titchener enables seamless hybrid-cloud management. In the next section, we show how it can be readily deployed in practice with a simple prototype system built from existing cloud technologies.

## 4 A Hybrid Kubernetes Platform

In this section, we design a prototype system based on the architecture proposed in Section 2 to allow the breaking up of Kubernetes (K8s) based data pipelines into partitions running seamlessly across public and private clouds.

Assuming the reader's familiarity with K8s abstractions: Resource, Pod, Service, Deployment and Custom Resource Definition (CRD), for convenience, we define some additional terminology here. Let $S = \{s_i\}$, $P = \{p_j\}$ be the set of services and pods in a K8s application. Then the application can be compactly described by a Pod-Service dependency function $f : P \times S \to [0, 1]$ where $f[p, s]$ is 1 if a pod $p$ needs to access a service $s$. $P(s)$ is the set of pods which need to access a service $s$.

### 4.1 System Design

We define a hybrid K8s platform as an interconnected set of K8s clusters where a public cloud hosted master Kubernetes cluster is connected to multiple private clusters.

**Our goal**: To enable seamless partitioning of any application across a hybrid K8s platform comprised of $C = \{c_i\}$ clusters where each disjoint partition $c_i$ represents the set of pods allotted to the $i^{th}$ cluster. For simplicity, we only allow partitioning schemes where all pods backing a service are allotted to the same partition i.e for each service $s$, $host\_cluster[s]$ is unique and points to the partition holding its pods. In this discussion, the application in question refers to the management plane of the pipeline. We assume the data-plane components are separately bootstrapped into each environment.

In our design, each K8s cluster runs Istio [18], a service mesh management software used to facilitate fine grained traffic routing and policy enforcement. Istio launches ingress and egress gateway services in each cluster (denoted by $igw[i]$ and $egw[i]$ for the $i^{th}$ cluster) which act as configurable proxies to route and forward traffic. Egress gateways serve as exit points to reach services running in external clusters while ingress gateways serve as entry points to intercept inbound traffic intended for services hosted on the cluster. In cluster $i$, connections to an external service $s_1$ are forced through its egress gateway $egw[i]$ at port $eport[i, s_1]$. Similarly traffic entering the cluster intended for a native-hosted service $s_2$ is forced through the ingress gateway $igw[i]$ at port $iport[i, s_2]$.

The public cloud component of the platform runs a master Kubernetes cluster, an application specific overwatch service and a pubsub message publisher which serves as an intelligent message dispatcher (Figure 1b). A typical setup of the platform involves an (1) initialization phase where necessary credentials are transferred to private networks and used to bootstrap control agent and data-plane components and (2) a configuration phase where the user uploads configuration to each control agent to establish connectivity among isolated application partitions and enforce global pod-service access constraints. For the sake of brevity we only describe steps involved in the configuration phase in greater detail here.

To initiate configuration, a user specifies the entire Pod-Service dependency graph and the partition allotted to each cluster as a Kubernetes CRD object. The message dispatcher broadcasts each CRD to every registered control agent. Each control agent processes received CRDs and extracts the application dependency graph and parti-tion allotted to it. It then independently executes the following three primitive operations (Algorithm 5):

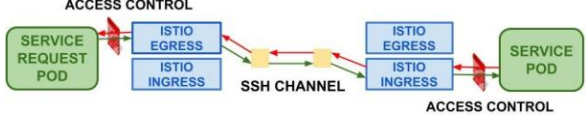

Figure 2: Flow of packets between a requesting Pod and a service backing pod split across two clusters.

```
def add_dns_entry(i, s):
    if host_cluster[s] ≠ i then
        IP[s] ← allot-unique-dummy-IP()
    end if
    return
```

**Algorithm 1:** Adds a DNS entry in the $i^{th}$ cluster for a service $s$ hosted by an external cluster.

```
def reserve_route(i, s, eport, iport):
    if host_cluster[s] ≠ i then
        eport[i, s] ← allot port(egw[i])
        Fwd(IP[s], Port[s] → egw[i], eport[i, s])
    else
        iport[i, s] ← allot port(igw[i])
        Fwd(igw[i], iport[i, s] → IP[s], Port[s])
    end if
    return
```

**Algorithm 2:** Reserves a route for a service $s$ in the $i^{th}$ cluster. If $s$ is hosted externally, Istio rules are created to forward traffic for $s$ to egress gateway. If not, traffic reaching the ingress gateway intended for $s$ is forwarded to its pods.

**(i) Service Discovery:** To allow pods in each partition to function seamlessly, all services they depend on should be discoverable by name. Since Kubernetes clusters only hold DNS entries for native services, additional dummy DNS entries are created by the control agent for services hosted in external partitions (Algorithm 1).

**(ii) Network connectivity:** To enable connectivity between pods across clusters, the control agent creates SSH channels to cloud hosted VMs and overlays local and remote port forwarding tunnels over them (Figure 1b). We claim that this is an effective technique to interlink management plane components if the volume of data exchange is typically small because (1) it avoids deployment and maintenance costs of expensive VPN hardware and (2) it requires minimal firewall modifications and can work in widely heterogeneous environments (like [14]) because no public IP addresses need to reserved. For each external (native) service, the control agent reserves a unique port on its egress (ingress) gateway and each private cluster creates SSH channels with local (remote) port forwarding to relay traffic between egress and ingress gateways across clusters. This is illustrated in Figure 2. Algorithms 2 and 4 briefly describe the associated operations.

**(iii) Access Control:** In this step, the control agent installs access control rules in its partition to block unauthorized service requests. For each pod, access to any service $s$ with $f(p, s) = 0$ is blocked. If $s$ is hosted outside its partition, the pod is prevented from reaching the corresponding mapped port $eport[i, s]$ on the relevant egress gateway $egw[i]$ (Algorithm 3).

```
def set_access_control(i, s, eport):
    block_all access(s)
    for each pod p ∈ (P(s) ∩ c_i) do
        if host_cluster[s] ≠ i then
            allow-access(p → egw[i], eport[i, s])
        else
            allow-access(p → IP[s], Port[s])
        end if
    end for
    if (host_cluster[s] == i) AND (P(s) ⊄ c_i)
    then
        allow-access(igw[i] → IP[s], Port[s])
    end if
    return
```

**Algorithm 3:** Sets access control for a service $s$ in the $i^{th}$ cluster. If $s$ is external, Istio is configured to allow access to egress gateway from select pods. If $s$ is reachable from external clusters, Istio is configured to allow traffic from ingress gateway to its service pods.

```
def create_channels(i, s, eport, iport):
    h ← host_cluster[s], m ← master_cluster
    if h == m then
        ch ← local_fwd(igw[m], iport[m, s])
        Fwd(egw[i], eport[i, s] → ch)
    else if h == i then
        ch ← remote_fwd(igw[i], iport[i, s])
        Fwd(ch → igw[i], iport[i, s])
    end if
    return
```

**Algorithm 4:** Interconnects the $i^{th}$ cluster and the master cluster. The functions `local_fwd` and `remote_fwd` create ssh channels to gateway VMs and add local and remote port forwarding to them. Additional Istio forwarding rules are created to link egress and ingress gateways with created ssh channels.

## 5 Hybrid-Cloud Composer Service

In this section we describe a specific use case of the hybrid K8s platform to extend a managed Apache Airflow service called Cloud Composer [16]. It is a workflow orchestration service that enables execution of pipelines specified as DAGs of data processing tasks. Each such task could directly process data or further trigger actions on data-plane components. Apache Airflow is a pertinent

```
def control_agent(f, i):
    m ← master cluster
    eport, iport ← {}
    for s ∈ S do
        add_dns_entry(i, s)
        reserve_route(i, s, eport, iport)
        set_access_control(i, s, eport)
    end for
    if i ≠ m then
        for s ∈ S do
            if (host_cluster[s] == m) then
                Estimate iport[m, s]
            end if
            create_channels(i, s, eport, iport)
        end for
    end if
    return
```

**Algorithm 5:** Control agent operations in $i^{th}$ cluster

example application where parts of data might reside on-premises and it might be desirable to run workers inside private environments to operate over them.

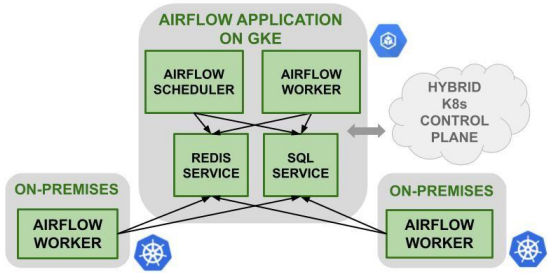

Figure 3: Hybrid-Cloud Composer: A cross cloud airflow driven workflow orchestration tool.

Cloud composer embeds Airflow components: web-server, scheduler, redis-broker, SQL database and workers into containers and services running on the Google Kubernetes Engine (GKE). In a typical setup, the scheduler continuously examines multiple DAGs and places tasks to be executed onto a message broker (e.g redis). Workers listen for new tasks to execute by monitoring the message broker and commit each finished task to an SQL database.

By modelling the airflow application as a simple pod-service dependency graph, we observe that an airflow worker interacts with message brokers and SQL database services throughout its lifetime. Thus, by simply leveraging the the proposed hybrid K8s platform (Section 4), message brokers and SQL database services running on GKE clusters can be seamlessly exposed to privately hosted airflow worker pods. Figure 3 illustrates the architecture of this cross cloud orchestration tool.

## 6 Related Work

Related works on this topic fall into two major categories: (1) API management and (2) Networking.

**API management:** Hybrid cloud applications are developed against API managed by different cloud vendors. It is difficult to create inter-operable services due to the lack of a unified abstraction layer [24] and significant variance among APIs available for similar tasks [25]. Open source libraries like [15, 19, 17] tackle some of these issues by presenting a single shared library to link against.

**Networking:** A Virtual Private Cloud (VPC) framework for inter-linking private data centers with public cloud over VPN connections was first proposed in [35]. In [21], the authors proposed a scalable VPN gateway architecture which can simultaneously track multiple high-bandwidth VPN connections. However, VPN infrastructure can be expensive to maintain for emerging edge-centric applications like MicroClouds and Cloudlets [27, 32, 26] where data transfer to the cloud is limited. Recent hybrid cloud proposals like Google Cloud's Anthos [14] (which offers hybrid Kubernetes extensions) can work in widely heterogeneous environments and make no assumptions about prior existence of VPN infrastructure.

## 7 Discussion

In the discussion thus far, we have largely treated data-plane partitions as black-box entities. These partitions may be loosely coupled or strongly interconnected with each other depending on whether there is cross-cloud data exchange. Enabling privacy-preserving and secure cross-cloud data exchange has been a topic of extensive study in the past [31, 30, 23]. But due to the associated complexity in implementing such schemes we envision that enterprise customers would likely prefer a edge-centric system where data exchange between private & public clouds is avoided altogether. In this regard, we believe our architecture and prototype implementation are already well-positioned to manage loosely-coupled data planes but they can also accommodate future proposals to monitor and enforce policies to support cross-cloud data exchanges.

## 8 Conclusion

In this paper we describe Titchener, an architecture for managing hybrid-cloud data processing pipelines. We vouched for a system comprising of independent and loosely coupled local control planes interacting with a highly available public cloud hosted master control plane. We implemented a hybrid Kubernetes platform based on the proposed architecture and addressed challenges pertaining to (1) service discovery (2) network connectivity for bi-directional data transfer and (3) access control enforcement across public and private clouds. We also pre-sented a specific use case of the platform to seamlessly extend a managed Apache Airflow service called Cloud Composer across multiple cloud environments.